\documentclass[a4paper,12pt]{article}

\usepackage[T2A]{fontenc}
\usepackage[english]{babel}
\usepackage{amsmath,amssymb}
\usepackage{cite}
\usepackage[unicode,linktocpage=true,plainpages=false,pdfpagelabels=false]{hyperref}
\usepackage{graphicx}
\usepackage{subfig}

\newcommand{\de}{\delta}
\newcommand{\De}{\Delta}
\newcommand{\ga}{\gamma}
\newcommand{\al}{\alpha}
\newcommand{\vk}{\varkappa}
\newcommand{\vp}{\varphi}
\newcommand{\la}{\lambda}
\newcommand{\m}{\mu}
\newcommand{\n}{\nu}
\newcommand{\ka}{\varkappa}
\newcommand{\ls}{\left(}
\newcommand{\rs}{\right)}
\newcommand{\ta}{\tau}
\newcommand{\dd}{\partial}

\newcommand{\disn}[2]{$$\displaylines{\refstepcounter{equation}%
            \label{#1}\hskip 1em minus 1em #2\hfilneg}$$}
\newcommand{\nom}{\hfil\hskip 1em minus 1em (\theequation)}
\newcommand{\no}{\hfil \hskip 1em minus 1em\phantom{(\theequation)}%
            \hfilneg\cr\hfilneg\hskip 1em minus 1em\hfil}
\newcommand{\ns}{\hfill\cr\hfill}
\begin{document}

\title{Weak field limit for embedding gravity}

\author{
S.~S.~Kuptsov$^{a,}$\footnote{E-mail: s1t2a3s4@yandex.ru},
M.~V.~Iof\/fe$^{b,}$\footnote{E-mail: ioffe000@gmail.com},
S.~N.~Manida$^{b,}$\footnote{E-mail: s.manida@spbu.ru},
S.~A.~Paston$^{b,}$\footnote{E-mail: pastonsergey@gmail.com}\\[0.5em]
{\it $^a$St.~Petersburg Department of Steklov Mathematical Institute of RAS,}\\
{\it St. Petersburg, Russia}\\[0.5em]
{\it $^b$ Saint Petersburg State University, St. Petersburg, Russia}
}
\date{\vskip 15mm}
\maketitle

\begin{abstract}
We study a perturbation theory for embedding gravity equations in a background for which corrections to the embedding function are linear with respect to corrections to the flat metric.
The arbitrariness remaining after solving the linearized field equations is fixed by an assumption that the solution is static in the second order.
A nonlinear differential equation is obtained, which makes it possible to find the gravitational potential for a spherically symmetric case if a background embedding is given.
An explicit form of a spherically symmetric background parameterized by one function of radius is proposed.
It is shown that this function can be chosen in such a way that
the gravitational potential is in a good agreement with the observed distribution of dark matter in a galactic halo.
\end{abstract}

\newpage

\section{Introduction}\label{razdvved}
Embedding gravity (also called embedding theory) \cite{regge} is a modified theory of gravity based on the idea of our spacetime as a 4D surface in a 10D flat space.
From a geometric point of view, this approach makes theory of gravity similar to the string theory.
In this case, space-time metric is considered as induced and has the form
\begin{equation}\label{g=dydy}
g_{\mu \nu} = (\partial _\mu y^a)(\partial _\nu y ^b) \eta _{ab},
\end{equation}
where $\eta _{ab}$ is a flat metric of the ambient space, and $y^a(x^\mu)$ is an embedding function defining the shape of the surface.
Here and further $\mu,\nu,\ldots=0,\ldots,3$ and $a,b,\ldots=0,\ldots,9$.
After the pioneering work \cite{regge}, the ideas of the embedding approach were repeatedly used in the works of various authors to describe gravity, including
the problem of
its quantization, see, for example, works
\cite{deser,tapia,maia89,frtap,estabrook1999,davkar,rojas09,statja25,faddeev,statja26}.

The embedding gravity equations of motion, called Regge-Teitelboim equations, turn out to be more general in comparison with Einstein's equations $G^{\mu\nu}=\varkappa\, T^{\mu\nu}$
and have the form:
\begin{equation}
	\left( G^{\mu\nu} -\varkappa\, T^{\mu\nu} \right) b^a_{\mu\nu} = 0,
	\label{RT}
\end{equation}
where $b^a_{\mu\nu}$ is a second fundamental form of a 4D surface defined by an embedding function $y^a(x^\mu)$.
It can be seen that all "Einsteinian"\ (i.~e.~satisfying Einstein's equations) solutions satisfy the Regge-Teitelboim equations, but the full set of solutions is not exhausted by them.
There may be \emph{extra} solutions for which $G^{\mu\nu} \neq \varkappa\, T^{\mu\nu} $, and nevertheless \eqref{RT} is satisfied.
Initially, this was seen as a disadvantage of the approach, since embedding gravity was understood as some reformulation of GR,
potentially more convenient for quantization due to the presence of a flat ambient space. Therefore, it was proposed to additionally introduce Einstein constraints into the theory, making it truly equivalent to GR \cite{regge,statja18}.
However, nowadays the presence of extra solutions in the theory can be considered as an advantage.

If we assume that just
the Regge-Teitelboim equations
describe the gravitational interaction, then from the point of view of
describing observations
in terms
of the usual GR, extra solutions will manifest themselves as an additional contribution to the right side of Einstein's equations.
It can be interpreted as a presence of some additional fictitious \emph{embedding matter}, which has nothing to do with ordinary matter.
This contribution depends solely on the gravitational variables (which in embedding theory are $y^a(x^\mu)$), and its appearance is connected only with an attempt to reformulate a new theory in the old language.
Since direct detection of dark matter does not currently yield results \cite{1509.08767,1604.00014},
it is of interest to try to treat the \emph{embedding matter} of the embedding theory as a \emph{dark matter} (and possibly as a dark energy).
Then the effects associated with dark matter and energy turn out to be purely gravitational.
In this way, observational problems that have no explanation within the framework of GR can be solved.
It is known the Regge-Teitelboim equations \eqref{RT} can be rewritten as a pair of equations
 \disn{sp2}{
G^{\m\n}=\ka \ls T^{\m\n}+\ta^{\m\n}\rs,
\nom}\vskip -2em
 \disn{sp2.1}{
\ta^{\mu\nu} b^a_{\mu\nu} = 0,
\nom}
i.~e.~in the form of a set of Einstein's equations with the contribution of an energy-momentum tensor $\ta^{\m\n}$ of fictitious embedding matter and its equations of motion.
It was first pointed out  in \cite{pavsic85}.
It should be noted that there are also modified theories of gravity alternative to embedding gravity, the equations of motion of which can be written in the form of Einstein's equations supplemented by other equations.
The most famous is mimetic gravity \cite{mukhanov,Golovnev201439}.

In order to obtain properties of the embedding matter generated by embedding gravity (and, therefore, to understand whether they are similar to the properties of dark matter), it is necessary to investigate solutions of the equations \eqref{RT}.
In general, due to their nonlinearity, this is too difficult a mathematical problem.
And so, we will limit ourselves to the most physically interesting
case of weak gravity, when the metric $g_{\m\n}$ is close to the flat metric $\eta_{\m\n}$.
In this case
the problem arises to choose
a background value $\bar y^a(x^\m)$ of the embedding function, which would correspond to the flat metric.
The simplest choice in the form of $\bar y^a(x^\m)$ which defines a 4D plane in the ambient space
turns out to be unsuitable, since in such a background the Regge-Teitelboim equations \eqref{RT} are not linearized
\cite{deser}.
To linearize the equations, it is necessary to choose "unfolded"\ embedding \cite{statja71} as the background, which means that the second fundamental form of the surface is nondegenerate in some sense.

In this paper we will use unfolded embedding of the Minkowski metric, which is
the product of a timelike line $y^I=const$ (we use indexes $I,K,\ldots=1,\ldots,9$; $i,k,\ldots=1,2,3$)
on 9D unfolded embedding $\bar{y}^I\left(x^i\right)$ of the euclidean 3D metric, i.~e.
 \disn{fon}{
	\bar{y}^a =
	\begin{pmatrix}
		x^0\\\bar{y}^I\left( x^i \right)
	\end{pmatrix},
\qquad
	\partial_i \bar{y}^I \partial_k \bar{y}^I = \de_{ik}.
\nom}
Note that with such a choice of background embedding, a nonrelativistic motion of embedding matter is possible \cite{statja68}.

The purpose of this work is to study the Regge-Teitelboim equations \eqref{RT} linearized in the background of \eqref{fon}.
We will look for a solution that corresponds to a galaxy which rotates so slowly that effect of the rotation can be neglected, i.~e. a static and spherically symmetric on average distribution of ordinary matter.
At the same time, we will assume that the metric \eqref{g=dydy} is also static (and spherically symmetric), which corresponds to the time independence of the
value $\ta^{\m\n}$ describing embedding matter.
The resulting solution determines the dependence of the gravitational potential on the distance to the center of the galaxy and it can be compared with observations
of the rotation curves of galaxies.

In sections~\ref{razdlin} and \ref{razdresh}, we obtain linearized equations and find their solution.
In section~\ref{razdsled}, the influence of the assumption of the exact static nature of the solution on its behavior in a linear approximation is investigated.
In section~\ref{razdsfer}, we study how the problem is simplified in the case of spherical symmetry.
We propose an explicit form of a spherically symmetric background embedding in section~\ref{razdvlog}.
In section~\ref{razdobr}, we study the possibility of choosing  this embedding in such a way that the corresponding gravitational potential is in agreement with the observed rotation curves of galaxies.

\section{Linearization of the Regge-Teitelboim equations}\label{razdlin}
Let us recall some formulas of the embedding theory.
All convolutions by Latin indexes are carried out using the flat metric of the ambient space $\eta_{a b}$.
The induced metric is expressed in terms of the embedding function by the formula \eqref{g=dydy}.
We will use the space-time signature $(-,+,+,+)$.
Note that the signature is changed by changing the sign of $\eta_{a b}$, and the induced metric changes the sign as a consequence.
The second fundamental form of a 4D surface is expressed in terms of the covariant derivative $D_\m$ of the embedding function consistent with the metric or through the projector $\Pi_{\bot}{}^a_b$ on a subspace transverse to the surface (see, for example, \cite{statja18}):
\begin{equation}
	b^a_{\mu\nu} = D_\mu \partial_\nu y^a = \Pi_{\bot}{}^a_b \partial_\mu \partial_\nu y^b.
	\label{b}
\end{equation}

We will mark with a line the values corresponding to the background embedding function $\bar{y}{}^a$ \eqref{fon}, for example
$\bar{b}{}^a_{\mu\nu}$ is the second fundamental form of the background surface.
We will raise and lower the 4D indexes of values with a line using the background metric.
Since it is flat, the background connection is zero, and the covariant derivative in \eqref{b}
is reduced to the usual one.
The second derivatives of $\bar{y}{}^0$ are zero, as are the derivatives by $x^0$ of $\bar{y}{}^I$, hence the nonzero components of $\bar{b}{}^a_{\mu\nu}$ are:
\begin{equation}
	\bar{b}{}^I_{ij} = \partial_i \partial_j \bar{y}^I.
	\label{b^I_ij}
\end{equation}
By the index $I$, this value
is transverse to the 3D surface in 9D space described by the embedding function $\bar{y}{}^I(x^i)$. Thus, the index $I$ at each point effectively runs through 6 values, exactly like the multi-index $\{ij\}$. Therefore, $\bar{b}{}^I_{ij}$ can be represented as a $6\times6$ matrix.
Since this matrix is nonsingular for "unfolded"\ embeddings (see details in \cite{statja71}), you can introduce symmetric by $l,m$ and transverse by $I$ value $\bar{\alpha}_I^{lm}$, which is inverse to $\bar{b}{}^I_{ij}$ in the matrix sense:
\begin{equation}
	\bar{\alpha}_I^{lm} \bar{b}{}^I_{ij} = \frac{1}{2} \left( \de^l_i \de^m_j +  \de^l_j \de^m_i \right).
	\label{alpha}
\end{equation}

We will look for solutions of the equations in the form of the system \eqref{sp2},\eqref{sp2.1} as a perturbation theory series in degrees of the gravitational constant $\vk$ in the background \eqref{fon}:
\begin{equation}\label{sp4}
y^a = \bar{y}^a  + \vk\overset{_{(1)}}{y}{}^a+\vk^2\overset{_{(2)}}{y}{}^a+\ldots.
\end{equation}
We will write similar expansions in a series of $\vk$ for any other quantities, including $\tau^{\mu\nu}$ and energy-momentum tensor of ordinary matter $T^{\m\n}$.
Substituting expansions into the system \eqref{sp2}, \eqref{sp2.1}, in the first non-vanishing order we get
 \disn{sp5}{
\overset{_{(1)}}G{}^{\m\n}= \bar T^{\m\n}+\bar\ta^{\m\n},
\nom}\vskip -2em
 \disn{sp6}{
\bar\ta^{ik} \bar b^I_{ik} = 0.
\nom}
Due to the nonsingularity of the \eqref{b^I_ij} mentioned above, the last equation corresponds to the condition
 \disn{sp5.1}{
\bar\ta^{ik}=0.
\nom}
Of the ten linearized Einstein equations \eqref{sp5}, only six (for $\m=i$, $\n=k$) turn out to be equations for the first correction $\overset{_{(1)}}y{}^a$ to the embedding function, and hence for the first correction $\overset{_{(1)}}g{}_{\m\n}$ to the metric, as follows from \eqref{g=dydy}.
The remaining four equations define the components of $\bar\ta^{0\m}$ and therefore do not restrict the embedding function.

It should be noted that only the contributions transverse to the background surface in $\overset{_{(1)}}y{}^a$ change the shape of the surface, so that only six components of $\overset{_{(1)}}y{}^a$ make a physical sense.
The others four components correspond to the longitudinal deformations of the surface, which are equivalent to diffeomorphisms.
Therefore, in order to find all 10 components of $\overset{_{(1)}}y{}^a$ uniquely, it is necessary to impose some coordinate conditions.

\section{Solving of the linearized equations}\label{razdresh}
Assume that at least in the zero order the distribution of ordinary matter is static, i.~e.~$\dd_0\bar T^{\m\n}=0$.
We will look for solutions of the linearized equations \eqref{sp5} that correspond
to the static addition to the metric, i.~e.~$\dd_0\overset{_{(1)}}g{}_{\m\n}=0$.
This corresponds to the fact that
in the main order the distribution of embedding matter will also be static, i.~e.~$\dd_0\bar\ta^{\m\n}=0$.

Instead of solving only six out of ten equations \eqref{sp5} (which can be done and gives the same result), it is convenient to solve all ten equations, treating the quantities of $\bar\ta^{0\m}$ as independent variables describing embedding matter.
Then the linearized Regge-Teitelboim equations are reduced simply to the linearized Einstein equations \eqref{sp5}.
We will assume that ordinary matter is dust-like (as can be done
for the matter in galaxies) and is at rest in zero order, then
\begin{equation}
\bar{T}_{\mu\nu} = \bar\rho(x^i) \de^0_\mu \de^0_\nu.
\end{equation}
The equations \eqref{sp5} can be considered as equations not for $\overset{_{(1)}}y{}^a$ directly,
but as equations for the correction to the metric $\overset{_{(1)}}g{}_{\m\n}$, from which $\overset{_{(1)}}y{}^a$ is then found by solving the embedding equation \eqref{g=dydy}.

As is well known, the equations for $\overset{_{(1)}}g{}_{\m\n}$ take a simple form
 \disn{sp6.1}{
\De \overset{_{(1)}}g{}_{00}=-(\bar\rho(x^i)+\bar\ta_{00}),\qquad
\De \overset{_{(1)}}g{}_{jk}=-(\bar\rho(x^i)+\bar\ta_{00})\de_{jk},
\nom}\vskip -2em
 \disn{sp6.2}{
\De \overset{_{(1)}}g{}_{0k}=-2\bar\ta_{0k}
\nom}
(where $\De$ is the Laplace operator)
if we choose harmonic coordinates and assume that $\overset{_{(1)}}g{}_{\m\n}$ does not depend on time. In addition, we will assume that $\overset{_{(1)}}g{}_{\m\n}$ is decreasing
at spatial infinity.
The conditions for the harmonic coordinates are reduced to
only one additional equation:
 \disn{sp7}{
\dd_k \overset{_{(1)}}g{}_{0k}=0.
\nom}
Introducing the gravitational potential $\vp(x^i)$ (notice that we do not include the gravitational constant $\vk$ in it for convenience) as a solution of the Poisson equation
\disn{sp7.1}{
\De\vp(x^i)=\frac{1}{2}(\bar\rho(x^i)+\bar\ta_{00})
\nom}
and taking into account
that the solutions is decreasing at spatial infinity, we obtain
 \disn{sp8}{
\overset{_{(1)}}g{}_{00}=-2\vp(x^i),\qquad
\overset{_{(1)}}g{}_{jk}=-2\vp(x^i)\de_{jk}.
\nom}

Now, we shall find the correction $\overset{_{(1)}}y{}^a$ to the embedding function which corresponds
to an arbitrary correction $\overset{_{(1)}}g{}_{\m\n}$ to the metric.
From the formula \eqref{g=dydy} in the first order by $\vk$ we have:
\begin{equation}
	\overset{_{(1)}}{g}{}_{\mu\nu} = (\partial_\mu \bar{y}^a)( \partial_\nu \overset{_{(1)}}{y}{}_a) + (\partial_\nu \bar{y}^a)( \partial_\mu \overset{_{(1)}}{y}{}_a ).
	\label{h=dydy}
\end{equation}
Let us introduce a parameterization for the correction to the embedding function as the sum of a tangent and orthogonal to the background components:
\begin{equation}\label{sp10}
	\overset{_{(1)}}{y}{}^a  = \overset{_{(1)}}{y}{}^\ga_\| \partial_\gamma \bar{y}^a  + \overset{_{(1)}}{y}{}_\bot^a .
\end{equation}
Notice that the structure of the background embedding \eqref{fon} leads to the relation $\overset{_{(1)}}{y}{}^0=\overset{_{(1)}}{y}{}^0_\|$.
Substitution of the representation \eqref{sp10} in \eqref{h=dydy} yields a system of differential equations for $\overset{_{(1)}}{y}{}^\ga_\|$ and $\overset{_{(1)}}{y}{}_\bot^a$:
 \disn{sp9}{
\eta_{\ga\n}\dd_\m \overset{_{(1)}}{y}{}^\ga_\| +\eta_{\ga\m}\dd_\n \overset{_{(1)}}{y}{}^\ga_\| -2\overset{_{(1)}}{y}{}_\bot^I \bar b^I_{\m\n}
=\overset{_{(1)}}{g}{}_{\mu\nu}.
\nom}
Writing out its components, we obtain a system of equations:
 \disn{sp9.1}{
-2\dd_0 \overset{_{(1)}}{y}{}^0=\overset{_{(1)}}{g}{}_{00},\no
-\dd_k \overset{_{(1)}}{y}{}^0+\dd_0 \overset{_{(1)}}{y}{}^k_\|=\overset{_{(1)}}{g}{}_{0k},\no
\dd_i \overset{_{(1)}}{y}{}^j_\|+\dd_j \overset{_{(1)}}{y}{}^i_\|-2\overset{_{(1)}}{y}{}_\bot^I \bar b^I_{ij}
=\overset{_{(1)}}{g}{}_{ij}.
\nom}
It can be solved by integrating the equations over time and taking into account the static $\overset{_{(1)}}{g}{}_{\mu\nu}$.
Denoting the integration constants as $\widetilde{y}^a(x^i)$, we write the answer as:
 \disn{sp11}{
\overset{_{(1)}}{y}{}^0 = -\dfrac{1}{2} \overset{_{(1)}}{g}{}_{00}x^0 + \widetilde{y}^0(x^i)=
\vp(x^i)x^0 + \widetilde{y}{}^0( x^i),\no
	\overset{_{(1)}}{y}{}^k_\| = \int\!\left( \overset{_{(1)}}{g}{}_{0k} + \partial_k\overset{_{(1)}}{y}{}^0 \right)dx^0 + \widetilde{y}^k(x^i),\no
	\overset{_{(1)}}{y}{}_\bot^I = \dfrac{1}{2} \bar{\al}^{kj} _I \left( -\overset{_{(1)}}{g}{}_{kj} + 2\, \partial_k \overset{_{(1)}}{y}{}^j_\| \right)
= \bar{\al}^{kj} _I \left( \vp(x^i) \de_{kj} + \partial_k \overset{_{(1)}}{y}{}^j_\| \right),
\nom}
where \eqref{sp8} was also used.

Substituting these expressions into \eqref{sp10}, we finally obtain
 \disn{sp12}{
\overset{_{(1)}}{y}{}^a=
\de^a_0\ls\vp(x^i)x^0 + \widetilde{y}{}^0( x^i)\rs+\ns+
\de^a_I\ls\ls\int\!\left( \overset{_{(1)}}{g}{}_{0k} + \partial_k\overset{_{(1)}}{y}{}^0 \right)dx^0 + \widetilde{y}^k(x^i)\rs\dd_k\bar y^I+
\bar{\al}^{kj} _I \left( \vp(x^i) \de_{kj} + \partial_k \overset{_{(1)}}{y}{}^j_\| \right)
\rs.
\nom}
This expression gives the solution of the linearized Regge-Teitelboim equations.
It is parameterized by the gravitational potential $\vp(x^i)$, which by equation \eqref{sp7.1}
is associated with the zero order $\bar\ta_{00}$ of the density of embedding matter, and also by the
components of the metric $\overset{_{(1)}}{g}{}_{0k}$, which by equation \eqref{sp6.2}
are related to the zero order $\bar\ta_{0k}$ of the flux density of this matter.
Notice that due to \eqref{sp6.2} and \eqref{sp7}  the components of $\bar\ta_{0k}$ obey
the condition $\dd_k\bar\ta_{0k}=0$ (it has the meaning of the embedding matter conservation condition in the case under consideration),
so only two of these three components can be considered as independent.

It is interesting to compare this result with the one obtained in \cite{statja68} in the framework of the nonrelativistic approximation.
For this purpose, let us write down the time derivative of
the solution \eqref{sp12}:
 \disn{sp13}{
\dd_0\overset{_{(1)}}{y}{}^0=\vp(x^i),\no
\dd_0\overset{_{(1)}}{y}{}^I=
\ls \overset{_{(1)}}{g}{}_{0k} + \partial_k\overset{_{(1)}}{y}{}^0\rs\dd_k\bar y^I+
\bar{\al}^{ij} _I
\dd_i \ls\overset{_{(1)}}{g}{}_{0j} + \partial_j\overset{_{(1)}}{y}{}^0\rs.
\nom}
The second equation reproduces equation (62) from \cite{statja68}, taking into account the assumption $\overset{_{(1)}}{g}{}_{0j}=0$ made there.
However, the first equation differs from equation (61) of \cite{statja68} by the absence of a quadratic contribution.
Of course, this is not surprising since the solution of \eqref{sp12} is obtained in the linear approximation, and hence with rejection of the quadratic contributions.
But when analyzing the nonrelativistic limit in \cite{statja68}, it was taken into account that for the
large intervals of time $x^0$
(when the time $x^0$ is replaced by the nonrelativistic time $t=x^0/c$, where $c$ is the speed of light) quadratic corrections can give a contribution comparable to the linear ones.

\section{Accounting for equations in the next order}\label{razdsled}
The solution obtained above in the first order of perturbation theory depends on the choice of distribution of embedding matter in zero order characterized by $\bar\ta_{0\m}$.
Let us find restrictions on this distribution assuming that both $\ta^{\m\n}$ characterizing embedding matter and the energy-momentum tensor of ordinary matter $T^{\m\n}$ are static not only in the zero approximation, but also exactly.
In particular, it follows  that the metric is also exactly static.
When considering a galaxy, it means that the configuration of both ordinary and embedding matter at some point passed into a stationary state. We will investigate just this particular case.

Let us write down the equation \eqref{sp2.1} in the next order in $\vk$ compared to the already studied equation \eqref{sp6}, and also use it:
 \disn{sp14}{
\overset{_{(1)}}{\ta}{}^{jk} \bar b^a_{jk}+\bar\ta^{00}\overset{_{(1)}}{b}{}^a_{00}+2\bar\ta^{0k}\overset{_{(1)}}{b}{}^a_{0k}
= 0.
\nom}
Notice that due to \eqref{b} and the properties of the background embedding \eqref{fon} we have
$\overset{_{(1)}}{b}{}^a_{0\m}=\bar\Pi_{\bot}{}^a_b \partial_0 \partial_\mu \overset{_{(1)}}{y}{}^b$.
This allows us to express $\overset{_{(1)}}{\ta}{}^{jk}$ from \eqref{sp14} in the form:
 \disn{RT_2nd}{
\overset{_{(1)}}{\ta}{}^{jk}=
- \left( \bar{\ta}{}^{0 0} \partial_0 \partial_0 \overset{_{(1)}}{y}{}^I + 2 \bar{\ta}{}^{i0} \partial_i \partial_0 \overset{_{(1)}}{y}{}^I \right) \bar{\al}^{jk}_I.
\nom}
Therefore, the assumed time independence of $\ta^{\m\n}$
leads to additional restrictions on the quantities in the right-hand side.
Time derivative of the equation \eqref{RT_2nd} together with  \eqref{sp13} and the condition $\partial_0 \overset{_{(1)}}{\ta}{}^{jk} = 0$ provide the equation
 \disn{statogr_ij}{
\bar{\ta}{}^{i0} \left( \left( \de^j_i\de^k_l + \de^j_l \de^k_i \right) \partial_l \vp + 2\, \partial_i \left( \bar{\al}^{lm}_I \partial_l \partial_m \vp \right) \bar{\al}^{jk}_I \right) = 0.
\nom}

Then we can also write down the equation \eqref{sp2} in the next order and it will provide the next
correction $\overset{_{(2)}}{g}{}_{\m\n}$ to the metric.
However, it is possible to write down a consequence of this equation in which the determined
correction will not enter.
This is the corresponding order of the equation $D_\m \ta^{\m\n}=0$, reflecting the conservation (with covariant corrections) of embedding matter.
In the lowest order, this equation has the form  $\dd_\m \bar\ta^{\m\n}=0$ and is satisfied given \eqref{sp5.1} and \eqref{sp6.2},\eqref{sp7}.
In the next order it looks like
 \disn{sp15}{
\dd_\m \overset{_{(1)}}{\ta}{}^{\m\n}+\overset{_{(1)}}{\Gamma}{}^\m_{\m\ga}\bar\ta^{\ga\n}+\overset{_{(1)}}{\Gamma}{}^\n_{\m\ga}\bar\ta^{\m\ga}
=0.
\nom}
Taking $\n=k$ in this equation and using the equation \eqref{RT_2nd}, as well as the assumed time independence of $\ta^{\m\n}$ in all orders, from the condition $\partial_0 \overset{_{(1)}}{\ta}{}^{0k} = 0$ we get an equation
 \disn{statogr_0k}{
\partial_j \left( \left( \bar{\ta}{}^{00} \partial_0 \partial_0 \overset{_{(1)}}{y}{}^I + 2 \bar{\ta}{}^{i0} \partial_i\partial_0 \overset{_{(1)}}{y}{}^I  \right) \bar{\al}^{jk}_I \right) -\ns-
\bar{\ta}{}^{0k} \eta^{\m\al} (\partial_\al \bar{y}_a)  \partial_\mu \partial_0 \overset{_{(1)}}{y}{}^a
	- \bar{\ta}{}^{00} (\partial_k \bar{y}_a) \partial_0 \partial_0 \overset{_{(1)}}{y}{}^a  -
 2\bar{\ta}{}^{j0} (\partial_k \bar{y}_a) \partial_j \partial_0 \overset{_{(1)}}{y}{}^a=0,
\nom}
where \eqref{sp5.1} is used, as well as an expression for connectivity (see, for example, \cite{statja18})
\disn{sp16}{
\Gamma^\n_{\m\ga}=g^{\n\al}(\dd_\al y_a)\dd_\m\dd_\ga y^a,
\nom}
 from which it can be found that
 \disn{sp17}{
\overset{_{(1)}}{\Gamma}{}^\n_{\m0}=\eta^{\n\al}(\dd_\al \bar y_a)\dd_\m\dd_0 \overset{_{(1)}}{y}{}^a.
\nom}
Using in \eqref{statogr_0k} the formulas \eqref{sp13}, we obtain an equation that, together with \eqref{statogr_ij}, restricts
the choice of the background embedding $\bar y^a$ and $\bar\ta^{0\m}$, which parameterize the solution of the linear approximation found in the previous section.
Notice that from the remaining equation \eqref{sp15} at $\n=0$ the variable $\overset{_{(1)}}{\ta}{}^{k0}$ is not excluded, so this equation does not provide a new restrictions on the above quantities.

The background embedding of the form \eqref{fon} is parameterized by three functions, since nine components of $\bar y^I(x^i)$
are imposed by six conditions that the metric of the surface described by this function is flat Euclidean.
At first sight, the resulting number of equations for the background and $\bar\ta^{0\m}$ seems too large -- six equations \eqref{statogr_ij} and three equations \eqref{statogr_0k} for six unknowns
(recall that, as noted after \eqref{sp12}, only two of the three components $\bar\ta^{0k}$ are independent).
However, it is easy to see that nevertheless a solution exists, since the six equations \eqref{statogr_ij} can be satisfied by choosing three quantities $\bar\ta^{0k}$ as
 \disn{sp17.1}{
\bar\ta^{0k}=0.
\nom}
We will study the solutions corresponding just to such a choice.
Physically it means that in the zero approximation embedding matter is at rest.
We obtain that using the background \eqref{fon} leads to the fact that the time independence $\bar\ta^{\m\n}$ actually leads to \eqref{sp17.1} (otherwise there are fewer unknowns than equations), i.~e.~to the embedding matter at rest in the zero approximation.
Notice that in the analysis of the nonrelativistic limit in \cite{statja68} such a background was found when discussing the nonrelativistic character of motion of embedding matter.
Taking into account \eqref{sp6.2} and assuming decreasing corrections to the metric at 3D infinity, it follows from \eqref{sp17.1} that $\overset{_{(1)}}{g}{}_{0k}=0$.

Given that $\bar\ta^{0k}=0$, the remaining equations \eqref{statogr_0k} are simplified.
Using \eqref{sp13} we can now rewrite them as
 \disn{sp18}{
\partial_j \left( (2\De\vp-\bar\rho)\bar{\al}^{jk}_I \bar{\al}^{lm}_I \dd_l\dd_m\vp\right) -
(2\De\vp-\bar\rho) \dd_k\vp=0,
\nom}
where $\bar{\ta}{}^{00}$ was expressed from \eqref{sp7.1}.

These are three equations for four quantities, namely, three that parameterize the background embedding function $\bar y^I(x^i)$
and one describing the density of embedding matter $\bar\ta^{00}$.
Thus, it turns out that the linearized Regge-Teitelboim equations, supplemented by
the condition of exact statics of solutions, leave an arbitrariness in the choice of one function of coordinates $x^i$.
The physical meaning of this arbitrariness lies in the presence of arbitrariness in the general case when specifying the distribution of embedding matter at the initial moment of time.

\section{The case of spherical symmetry}\label{razdsfer}
Let us consider the situation when ordinary matter is distributed in a spherically symmetric way.
Then the solution can also be sought in the class of spherically symmetric functions.
In this case the gravitational potential $\vp$ will depend only on the radial coordinate $r=\sqrt{x^i x^i}$, as well as the distribution density of ordinary matter $\bar\rho$.

The question of how to formulate a condition of any kind of symmetry for an embedding function is rather non-trivial, see, for example, \cite{statja27}. Since the background embedding function $\bar y^I(x^i)$ enters the equation \eqref{sp18} only as a quantity
\begin{equation}
	A^{lm,jk} = \bar{\al}_I^{lm} \bar{\al}^{jk}_I,
	\label{A}
\end{equation}
it is possible to impose symmetry constraints directly on this.
Taking into account all index permutation symmetries, in the case of spherical symmetry the quantity $A$ can be written in the form
 \disn{A_sfersim}{
	A^{lm,jk}  = f_1(r) \de^{lm}\de^{jk} + \dfrac{1}{2} f_2(r) \left( \de^{lj}\de^{mk} + \de^{lk}\de^{mj} \right) + \dfrac{1}{2r^2} f_3(r) \left( \de^{lm} x^j x^k + \de^{jk} x^l x^m \right) + \ns
	+ \dfrac{1}{4r^2} f_4(r) \left( \de^{lj} x^m x^k +\de^{lk} x^m x^j +\de^{jm} x^k x^l +\de^{mk} x^j x^l \right) + \dfrac{1}{r^4} f_5(r) x^l x^m x^j x^k.
\nom}
Here the functions $f_1...f_5$ are characteristics of the particular background embedding used.
If we substitute this representation into \eqref{sp18}, a common multiplier $ x^k $ will appear, and the three equations will turn into one scalar equation.
Introducing notations for different combinations of functions $ f_1...f_5 $,
the resulting equation can be written in the following form:
\begin{equation}
	\left( \left( \vp'' + \dfrac{2}{r} \vp' - \dfrac{1}{2} \bar\rho \right) \left( \vp'' F_1 + \dfrac{1}{r} \vp' F_2 \right) \right)' + \dfrac{2}{r}\left( \vp'' + \dfrac{2}{r} \vp' - \dfrac{1}{2} \bar\rho \right) \left( \vp'' F_3 + \dfrac{1}{r} \vp' F_4 \right) = 0,
	\label{statorg_final}
\end{equation}
where
\begin{align}
	F_1 &= f_1 + f_2 + f_3 + f_4 + f_5,\\
	F_2 &= 2f_1 + f_3,\\
	F_3 &= f_2 + \dfrac{1}{2} f_3 + f_4 + f_5,\\
	F_4 &= - f_2 + f_3-\dfrac{r^2}{2} .
\end{align}

The equation \eqref{statorg_final} allows us to determine the gravitational potential $\vp(r)$ if the distribution of ordinary matter $\bar\rho(r)$ is known, as well as the background embedding function defining $f_1...f_5$.
Since \eqref{statorg_final} is a
nonlinear differential equation, in the general case its solution can only be sought numerically.
Previously, it is necessary to find all the functions $f_1...f_5$ included in the equation using the given embedding function, which requires first finding the value of $\bar{\al}_I^{lm}$.

Finding it directly by the formula \eqref{alpha} is a non-trivial task for specific embeddings.
Let us find a simpler way which allows us to find the functions $f_1...f_5$ included in the expansion of \eqref{A_sfersim} without
explicit knowledge of $\bar{\al}_I^{lm}$.
Notice that when writing this expansion, we took into account three types of symmetry by permutations of indexes following from the definition of \eqref{A}: inside the first pair, inside the second, and when replacing the first pair with the second.
As a consequence, we have five independent coefficients $ f_1...f_5 $.
But in fact there are fewer independent variables parameterizing $ A $.
To show this, we define quantity $ B $:
\begin{equation}
	B_{jk,is} = \bar{b}_{jk}^I\, \bar{b}_{is}^I=(\partial_j \partial_k \bar{y}^I) \partial_i \partial_s \bar{y}^I.
	\label{B}
\end{equation}
It is easy to see that this quantity is symmetric by permutation of any two indexes due to orthogonality of the first and second derivatives of $\bar{y}^I$:
\begin{equation}
	B_{jk,is} = (\partial_j \partial_k \bar{y}^I) \partial_i \partial_s \bar{y}^I = - (\partial_i \partial_j \partial_k \bar{y}^I) \partial_s \bar{y}^I = (\partial_i \partial_k \bar{y}^I) \partial_j \partial_s \bar{y}^I=B_{ik,js}.
\end{equation}
Hence, in the spherically symmetric case it can be parameterized not by five but only by three radius functions:
 \disn{B_sfersim}{
	B_{jk,is} = \dfrac{1}{3} g_1(r) \left( \de^{jk} \de^{is}  +\de^{js} \de^{ik}  +\de^{ji} \de^{sk}  \right) + \dfrac{1}{6r^2} g_2(r) \Big( \de^{jk} x^i x^s + \de^{ji} x^k x^s + \ns
	+ \de^{js} x^i x^k + \de^{ik} x^j x^s + \de^{is} x^j x^k + \de^{sk} x^j x^i \Big) + \dfrac{1}{r^4} g_3(r)\, x^j x^k x^i x^s.
\nom}
There is a connection between $A$ and $B$ that follows from \eqref{alpha}:
\begin{equation}
	A^{lm,jk} B_{jk,is} = \dfrac{1}{2} \left( \de^l_i \de^m_s + \de^l_s \de^m_i \right).
\end{equation}
It can be used to express five functions $ f_1...f_5 $ through three functions $ g_1...g_3 $:
\begin{align}
	f _1  &= \dfrac{-24 g_1^2 + (-24 g_2 - 36 g_3) g_1 + 3 g_2^2}{4 (20 g_1^2 + (20 g_2 + 24 g_3) g_1 - g_2^2) g_1}, \nonumber\\
	f _2  &= \dfrac{3}{2 g_1},  \nonumber\\
	f _3  &= \dfrac{(12 g_2 + 36 g_3) g_1 - 3 g_2^2}{2 (20 g_1^2 + (20 g_2 + 24 g_3) g_1 - g_2^2) g_1}, \label{f1..5(g1..3)}\\
	f _4  &= \dfrac{-3 g_2}{g_1 (2 g_1 + g_2)}, \nonumber\\
	f _5  &= \dfrac{-360 g_1^2 g_3 + (138 g_2^2 + 108 g_2 g_3) g_1 - 3 g_2^3}{4 (20 g_1^2 + (20 g_2 + 24 g_3) g_1 - g_2^2) g_1 (2 g_1 + g_2)}.\nonumber
\end{align}
As a result, to find the functions $f_1...f_5$ corresponding to a specific background embedding,
we first need to calculate the second fundamental form $\bar{b}{}^I_{ij}=\dd_i\dd_j \bar y^I$.
Then we have to determine according to \eqref{B} the quantity $B_{jk,is}$.
Comparing the result with the \eqref{B_sfersim} expansion, we can find the functions $g_1...g_3$, and then find the functions $f_1...f_5$ by \eqref{f1..5(g1..3)}.

\section{Explicit unfolded spherically symmetric embedding}\label{razdvlog}
Let us present an embedding $\bar{y}^I(x^i)$ of the flat Euclidean 3D metric into 9D Euclidean space,
which would be spherically symmetric in the sense discussed in \cite{statja27},
and would also be unfolded in terms of paper \cite{statja71}.
The latter means that for such a surface the second fundamental form \eqref{b^I_ij} is nonsingular as a $6\times6$ matrix,
see the text after \eqref{b^I_ij}.

We consider the basis $\la^A_{ij}$ in the space of symmetric traceless matrices,
numbered by the index $A$, which runs through the values $4,\ldots,8$.
We subject this basis to the orthonormality condition in the space of such matrices:
\begin{equation}
\la^A_{ij} \la^A_{lm} = \dfrac{1}{2} \left( \de_{il} \de_{jm} + \de_{im} \de_{jl} \right)-\dfrac{1}{3}\de_{ij} \de_{lm},\qquad
\la^A_{ij} \la^B_{ij} =\de^{AB}.
\end{equation}
We take the embedding function $\bar{y}^I(x^i)$ with components
\begin{align}
&\bar{y}^k(x^i) = f(r) x^k,\nonumber\\
&\bar{y}^A(x^i) = g(r) \la^A_{lm} x^l x^m,\nonumber\\
&\bar{y}^{9}(x^i) = h(r),\label{yxx}
\end{align}
it is parameterized by three radius functions.
One can check that this embedding function describes a surface with spherical symmetry. It means there is such a representation (generally speaking, reducible) of the group of 3D rotations under the action of which in the space $\bar{y}^I$ the surface transforms to itself (see \cite{statja27} for details).
The three independent functions $f, g, h$ are related to its irreducible parts, and the reducible representation as a whole can be described as "3+5+1"{}.

Let us select among all the embeddings of type \eqref{yxx} those that correspond to a flat metric.
Calculating for \eqref{yxx} the components of the induced metric, we obtain the equations
\begin{align}
	&f^2 + 2 r^2 g^2 = 1, \nonumber\\
	&\left( rf \right)'^2 + \dfrac{2}{3} \left( r^2 g \right)'^2 + h'^2 = 1.
	\label{yxx_g=eta}
\end{align}
Expressing from these equations the functions $g$ and $h$ through $f$ we obtain the desired spherically symmetric embedding of the flat Euclidean 3D metric into 9D Euclidean space.
It is parameterized by choosing one function of radius $f(r)$.
The fact that such an embedding is unfolded by a non-special choice of this function can be checked directly by computing the second fundamental form $\partial_i \partial_j \bar{y}^I$ \eqref{b^I_ij} and making sure that it is nonsingular.

Calculating this quantity and then the expression \eqref{B} constructed from it leads to the following expressions for the coefficients $g_1...g_3$ of its expansion \eqref{B_sfersim}:
\begin{align}
	g_1 &= 6 g^2, \\
	g_2 &= 6 \left( f'^2 + 2 r^2 g'^2 + 4r g g' \right), \\
	g_3 &= r^2 \left( \dfrac{h'}{r} \right)'^2 + r^2 f''^2 + 2r f' f'' - 3f'^2 + \dfrac{2}{3} r^4 g''^2 + 4 r^3 g' g'' + 4 r^2 g g'' - 2r^2 g'^2 - 4r g g' ,
\end{align}
where the functions $g$ and $h$ must be expressed through a single arbitrary function $f(r)$ according to \eqref{yxx_g=eta}.

\section{Inverse problem}\label{razdobr}
The equation \eqref{statorg_final} obtained above in the section~\ref{razdsfer}
allows one to search for the gravitational potential $\vp(r)$ if the distribution of ordinary matter $\bar\rho(r)$ and the background embedding function $\bar{y}^I(x^i)$ are known.
It is easy to see that among the solutions of this equation there is an Einsteinian solution for which
 \disn{sp19}{
\vp'' + \dfrac{2}{r} \vp' - \dfrac{1}{2} \bar\rho=0.
\nom}
However, there are other solutions too.
In the galactic halo region, where the contribution of $\bar\rho$ can be neglected, we have a Newtonian behavior of the gravitational potential $\vp\sim 1/r$ for the Einsteinian solution.
As we know, such a behavior does not agree well with the data of the rotation curves of galaxies.
To find other solutions in this area, it is sufficient to know the background embedding $\bar{y}^I(x^i)$.

The background embedding constructed in the previous section has ambiguity in the choice of one function $f(r)$.
There is no highlighted natural way to choose this function.
As a consequence there is also no possibility to uniquely find the gravitational potential $\vp(r)$ as a result of solving the equation \eqref{statorg_final}, since changing $f(r)$ will also change $\vp(r)$.
Therefore, we will focus on investigating the inverse problem by posing the question: can we find such a background embedding $\bar{y}^I(x^i)$ (i.~e.~such a function $f(r)$) so that the solution of the equation \eqref{statorg_final} corresponds to the gravitational potential in good agreement with galactic halo observations?

As a good fitting of observational data, we will take the Burkert profile \cite{burkert} of the dark matter density, see, for example, review \cite{2005.03520}:
 \disn{sp20}{
\rho_B(r)=\frac{\rho_0 r_0^3}{(r+r_0)(r^2+r_0^2)},
\nom}
where $r_0$ is the core radius, $\rho_0$ is the density at the center of the galaxy.
In order to find a behavior of the gravitational potential that agrees well with the observations,
one needs to solve the equation
 \disn{sp20.n1}{
\vp''+\frac{2}{r}\vp'=\frac{1}{2}(\bar\rho+\rho_B).
\nom}
The solution of this equation in the galactic halo has the form:
 \disn{sp20.n2}{
\vp(r)=\frac{M}{r}+\frac{\rho_0}{2}\ls\ls 1+ \frac{r_0}{r}\rs \ls \arctan\frac{r}{r_0}-\ln\ls 1+\frac{r}{r_0}\rs\rs+\frac{1}{2}\ls 1- \frac{r_0}{r}\rs\ln\ls 1+\frac{r^2}{r_0^2}\rs\rs,
\nom}
where $M$ is the mass of ordinary matter in the galaxy.
To avoid introducing this quantity as an additional parameter, we will further assume that the proportional to $M$ contribution to the gravitational potential in the halo region can be neglected compared with the contribution of dark matter.
Trivial substitution of function \eqref{sp20.n2} with $M=0$ into the equation \eqref{statorg_final} leads to a very complicated (so we do not show it here) third-order nonlinear differential equation for the unknown function $f(r)$ parameterizing the background embedding.
This equation fails to be solved analytically.
We have numerically solved the Cauchy problem for obtained equation, choosing different values of the initial data at the value of the radius $r=r_0/2$, which is usually about the visual size of the galaxy.
Recall that we solve the equation \eqref{statorg_final} with zero contribution of ordinary matter, so it is applicable only in the region of the galactic halo.

\begin{figure}[t]\centering
	\subfloat[]{\includegraphics[width=0.47\textwidth]{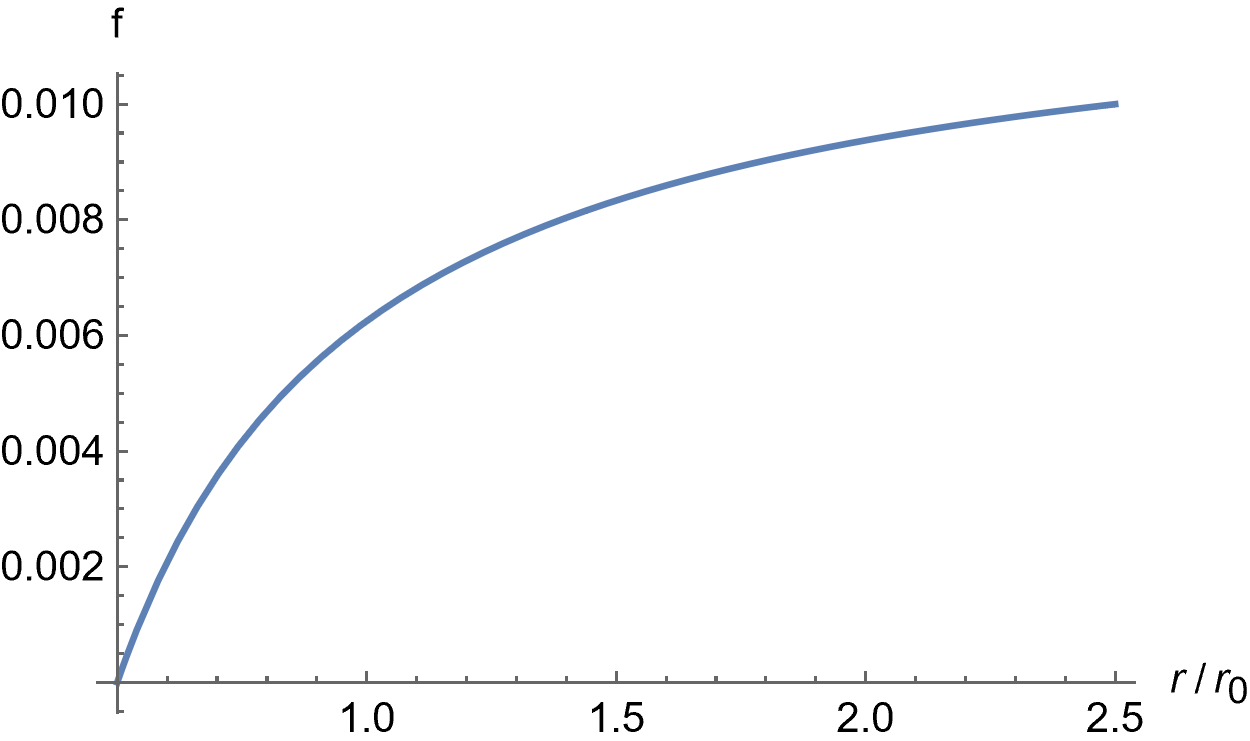}}\quad
	\subfloat[]{\includegraphics[width=0.47\textwidth]{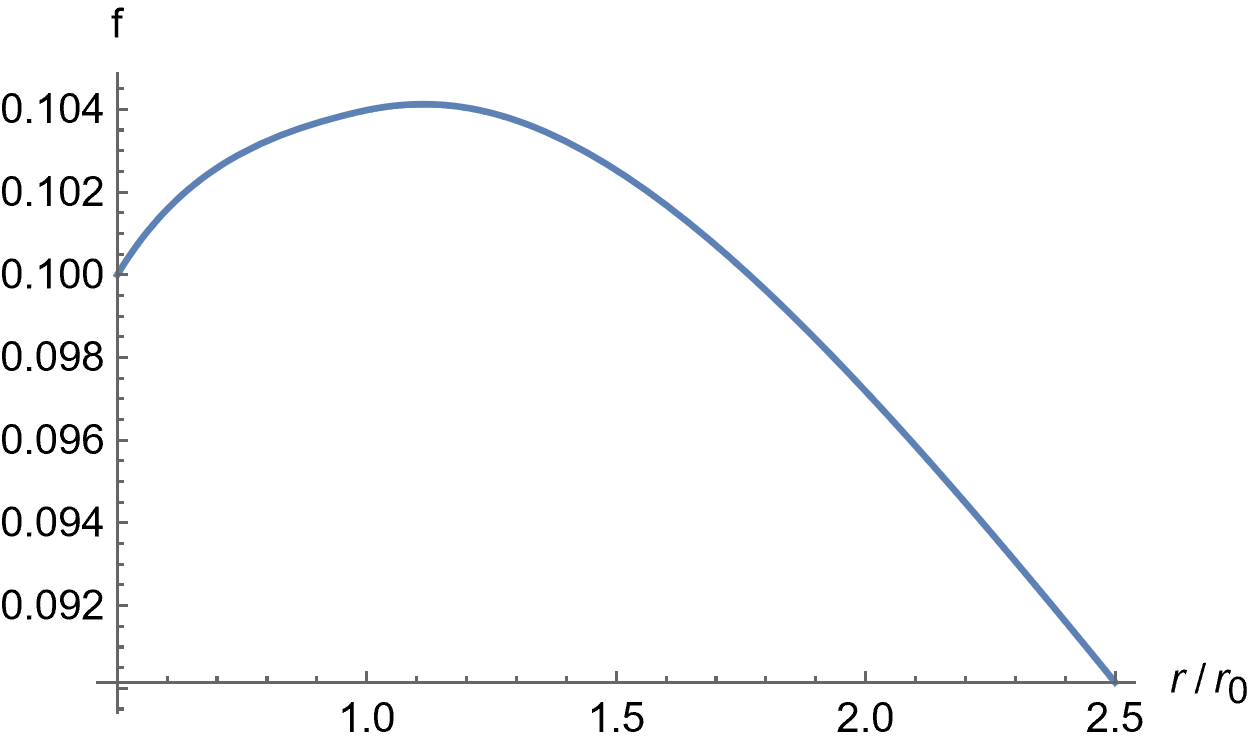}}
	\caption{Numerically constructed solutions with different initial data:\\ (a) $f=0$, $f'=0.025$, $f''=-0.1$; (b) $f=0.1$, $f'=0.02$, $f''=-0.1$.}
\end{figure}

A typical view of the obtained solutions is shown in Fig.~1.
The existence of such solutions shows the possibility to choose parameterizing the embedding \eqref{yxx} function $f(r)$ in a special way.
For this choice, equations of motion in the framework of linearization of the embedding theory in this background
lead to the same gravitational potential as the dark matter with Burkert profile \eqref{sp20} in the framework of ordinary GR.
Note here that with a significant change in the chosen values of the initial data, the algorithm used for the numerical solution of the nonlinear differential equation no longer made it possible to extend the solution into the region $r \gtrsim 0.95r_0$.
This can be explained by the fact that in the differential equation being solved, the coefficient at the major derivative (which itself is a compound quantity that also depends on the minor derivatives) may become too small, which prevents the correct behavior of the algorithm.

\section{Conclusion}
We considered the problem of constructing a perturbation theory for the Regge-Teitelboim equations \eqref{RT} for some background.
The background is chosen as \eqref{fon}, which corresponds to a 4D surface that is a direct product of the time-like straight line on the 9D unfolded embedding of the euclidean 3D metric.
This formulation of the problem is a special case of the weak gravitational field approximation in the embedding theory, in which the corrections to the embedding function are linear to the corrections to the metric, see ~\eqref{sp11}.
The analysis is simplified if the Regge-Teitelboim equations are rewritten as a set of Einstein equations \eqref{sp2} with the contribution
of fictitious embedding matter and the equations \eqref{sp2.1} describing the behavior of this matter.

Firstly, the Einstein equations are linearized in the usual way and solved using harmonic coordinates.
The result depends on the energy-momentum tensor of embedding matter specified arbitrarily in the lower order.
Further, we restrict ourselves to the class of solutions for which both ordinary matter and embedding matter are in a stationary state, which corresponds to the description of an already fully formed galaxy.
Taking this condition into account in the second order of perturbation theory leads to additional restrictions on the solutions found in the linear one.
As a result, the equation \eqref{statorg_final} arises in the case of spherical symmetry, which allows one to find the gravitational potential $\vp(r)$ if the background embedding is given.
Notice the nonlinearity of this equation, which is explained by the fact that it arises as a result of accounting for the static character of the metric in the second (following the linear) order.

It is possible to find a class of spherically symmetric surfaces \eqref{yxx}, \eqref{yxx_g=eta}, which have all necessary properties for their use as a background embedding.
These embeddings are parameterized by a single function $f(r)$.
It is shown that this function can be chosen so that the solution of the equation \eqref{statorg_final} leads to a gravitational potential that corresponds well (if we neglect deviations from spherical symmetry for real galaxies) to the observed distribution of dark matter in the galactic halo.

In our analysis we used the following assumptions.
We assume that ordinary matter has on average a static spherically symmetric distribution.
As a consequence, the energy-momentum tensor of ordinary matter $T^{\m\n}$ has exactly  spherically symmetric and static character (we use this fact in the first two orders of perturbation theory: for $\bar T^{\m\n}$ and $\overset{_{(1)}}{T}{}^{\m\n}$).
We further assume that the metric also has exactly  spherically symmetric and static character, and we use this for corrections to the metric $\overset{_{(1)}}{g}{}_{\m\n}$ and $\overset{_{(2)}}{g}{}_{\m\n}$.
Given the restrictions on $T^{\m\n}$, imposing such an assumptions on the metric is the same as imposing them on the energy-momentum tensor of embedding matter in the first two orders: for $\bar \tau^{\m\n}$ and $\overset{_{(1)}}{\tau}{}^{\m\n}$.
As a background embedding function $\bar y^a(x^\mu)$ which corresponds to the flat metric $\eta_{\m\n}$, we take the product \eqref{fon} of a timelike line on 9D unfolded and the spherically-symmetric embedding \eqref{yxx} of the euclidean 3D metric.
The assumptions made are internally consistent, but for real galaxies they are satisfied only approximately.
The strongest ones are the deviations from the spherical symmetry.

We discussed the behavior of the gravitational potential $\vp(r)$ only in the galactic halo.
In order to study the behavior of $\vp(r)$ inside the galaxy, where the influence of ordinary matter cannot be neglected, one must additionally find an explicit form of its distribution, which is beyond the scope of this paper.
At the same time, it should be emphasized that deviations of embedding gravity predictions from GR arise at scales where the energy-momentum tensor $\tau^{\m\n}$ of embedding matter becomes significant compared to the energy-momentum tensor $T^{\m\n}$ of ordinary matter.
We assume that at the scale of the galaxy these two quantities are comparable, but at smaller distances the influence of the embedding matter can be neglected and the laws of GR start to be satisfied with good accuracy.

\vskip 1em

{\bf Acknowledgements.}
The work of S.S.~Kuptsov was supported by Leonhard Euler International Mathematical Institute Grant No.~075-15-2019-1620.
The work of S.A.~Paston was supported by RFBR Grant No.~20-01-00081.

%\bibliographystyle{../../my3}
%\bibliography{../../paston-grav-e}
%\end{document}

\end{document}